\newcommand{\ta}{\ensuremath{\tau}~}
\newcommand{\ttp}{\ensuremath{\tau^{-}\to\pi^{-}\pi^{-}\pi^{+}\nu_{\tau}}}

\title{Kinematic reconstruction  of  $Z/H \rightarrow \tau\tau$ decay  in proton-proton collisions}

\author{
        Vladimir Cherepanov \\
                Institut Pluridisciplinaire Hubert Curien (IPHC)\\
        67037  Strasbourg, France \\
            \and
        Alexander Zotz\\
        III. Physikalisches Institut B, RWTH Aachen University \\
        D-52056 Aachen, Germany \\
}

\date{\today}

\documentclass[12pt]{article}

\usepackage{amsmath}
\usepackage{graphicx}
\usepackage{hyperref}

\begin{document}

\maketitle

\begin{abstract}
The knowledge of \ta lepton kinematic
and kinematic of the \ta pair in the decay $Z/H \rightarrow \tau\tau$ is essential for various analysis at LHC.
However, the reconstruction of the whole kinematic of the \ta decay is a challenging task, since in 
every \ta decay at least one neutrino is present in the final state which escapes detection.  In this paper  a 
kinematic technique (Global Event Fit) to estimate the momentum escaped with neutrinos and hence the full momentum of the \ta lepton pair is described. The algorithm is based on
iterative minimization  of the likelihood with  constrains  derived from  all available kinematic information on the decay.  The method requires 
 the direction of at least one \ta lepton to be well defined and therefore the method can be applied to the decays $Z/H \rightarrow \tau\tau \rightarrow X  + a_{1}\nu$
with $a_{1}$ resonance decaying into three charged pions.
\end{abstract}

\section{Introduction}\label{kinfitchap}

The momentum of the  \ta  lepton is not measuable in the detector due to the presence of at least one neutrino in its decay. In $pp$ or $p\bar{p}$ collisions 
the center of mass energy of the \ta pair is not known.  Instead only the imbalance of the transverse energy (missing transverse energy (MET)) 
is available, which represents the sum of neutrinos momentum in the event. However, for various analyzes, especially those that require the
knowledge of \ta spin state the reconstruction of the full \ta momentum is essentially important.
For this purpose the Gloabl Event Fit algorithm, described in this paper, can be used. The method consist in 
iterative search for the $\tau\tau$ kinematic that best satisfies the constraints that can be applied to $\tau\tau$ system
and the decay products of \ta leptons.

The mathematical framework used for the fit is based on notes by Paul Avery \cite{avery:cbx9172,avery:cbx9837}.
The general procedure is divided into three steps:

\begin{enumerate}
\item  The reconstruction of the primary vertex (PV) and the secondary vertex (SV) of the hadronically decaying \ta . %
\item  The calculation of the $\tau$ momentum in the decay \ttp. 
\item  The reconstruction of the momentum of both $\tau$ leptons by applying a kinematic fit with constraints on  the di- \ta system.
\end{enumerate}

Each step is discussed in details in the following sections.

\section{Reconstruction of the $\tau \rightarrow a_{1}\nu$ direction}

The large multiplicity of proton-proton collisions allows to reconstruct the primary vertex  -  the point of \ta pair production. In
order to reduce the bias on the vertex position created by the decay products of the \ta lepton the PV has to  be refitted  after removing the
visible decay products from the \ta leptons.


Assuming no flight length of the strongly decaying $a_1$ resonance the point of \ta decay  can be reconstructed 
by fitting three charged tracks from $a_{1}$ decay requiring them to originate from the common space point. 
With the knowledge of both vertices the flight direction of the \ta lepton is determined 
as:

\begin{equation}\label{eqtaudir}
\vec{n}_{\tau}  = \frac{\vec{SV} - \vec{PV}}{|\vec{SV} - \vec{PV}|}.
\end{equation}

\section{Calculation  of the \ta momentum in the decay  $\tau \rightarrow a_{1}\nu$} \label{step2}
 The momentum of the \ta lepton in the two-body $\tau \rightarrow 3\pi^{\pm} + \nu_{\tau}$ decay  can be calculated using energy and momentum conservation.
Assuming the neutrino to be massless, in the laboratory frame the following holds:

\begin{equation}\label{eq51}
(P_{\tau} - P_{a_{1}})^{2} = 0,
\end{equation}
where $P_{\tau}$ and $P_{a_{1}}$ are the four-momenta of the \ta lepton and the $a_{1}$ respectively. 
The \ta momentum reads then:

\begin{equation}\label{TauMom}
|\vec{p}_{\tau}| = \frac{(m^{2}_{a_{1}} + m^{2}_{\tau})|\vec{p}_{a_{1}}|\cos\theta_{GJ} \pm \sqrt{(m^{2}_{a_{1}} + \vec{p}^{2}_{a_{1}})[(m^{2}_{a_{1}} - m^{2}_{\tau})^{2} - 4m^{2}_{\tau}\vec{p}^{2}_{a_{1}}\sin^{2}\theta_{GJ}]} }{2(m^{2}_{a_{1}} + \vec{p}^{2}_{a_{1}}\sin^{2}\theta_{GJ})} .
\end{equation}

The Gottfried-Jackson angle $\theta_{GJ}$ is defined as the angle between the directions of the \ta lepton and the $a_{1}$ in the laboratory frame.  
For a given $\theta_{GJ}$ and momentum of the $a_{1}$ two values for the \ta momentum are possible. This ambiguity vanishes if the square root in Eq.~(\ref{TauMom}) is zero, it
happens if the Gottfried-Jackson angle $\theta_{GJ}$ reaches its  maximum allowed value $\theta^{max}_{GJ}$;

\begin{equation}\label{gjmax}
\begin{split}
&(m^{2}_{a_{1}} + \vec{p}^{2}_{a_{1}})[(m^{2}_{a_{1}} - m^{2}_{\tau})^{2} - 4m^{2}_{\tau}\vec{p}^{2}_{a_{1}}sin^{2}\theta_{GJ}] = 0; \hspace{0.2cm}  \\
&\theta^{max}_{GJ} = arcsin\frac{m^{2}_{\tau} - m^{2}_{a_{1}}}{2m_{\tau}|\vec{p}_{a1}|}  . \\
\end{split}
\end{equation}

An illustration of the kinematics is given in Fig.~\ref{AmbigRota} ({\textit{Left}). The solid curve shows the value of the \ta momentum as a 
function of $\theta_{GJ}$. 
For a given value of $\theta_{GJ}$ the two solutions of Eq.~(\ref{TauMom}) 
are denoted as positive and negative for the higher and lower value of the $\tau$ momentum respectively.  
The ambiguity point corresponds to the $\tau$ momentum at the maximum value of $\theta_{GJ} = \theta_{GJ}^{max}$.

\begin{figure}[]  
  \begin{center}
    \includegraphics[width=0.48\textwidth]{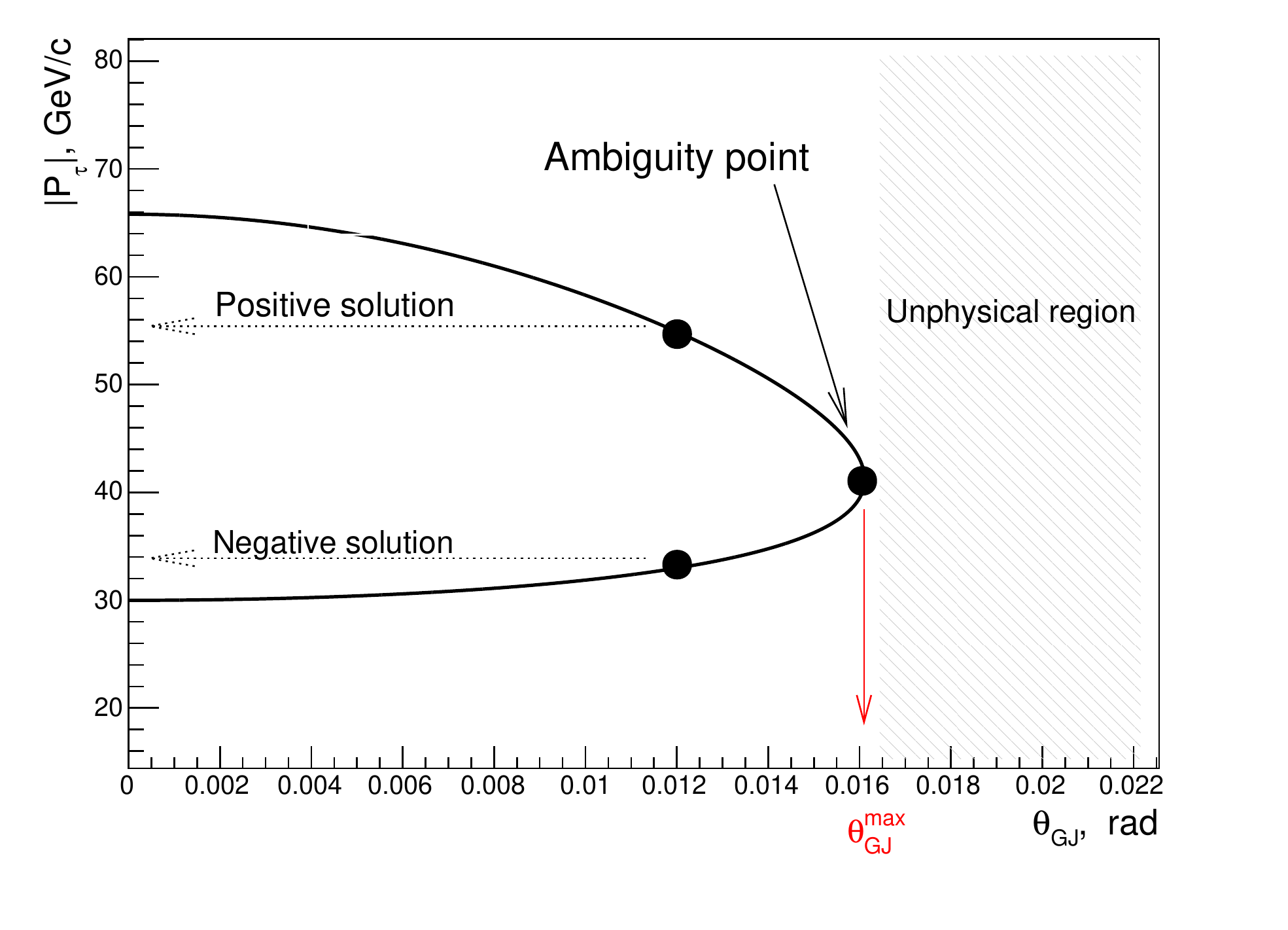}
    \includegraphics[width=0.35\textwidth]{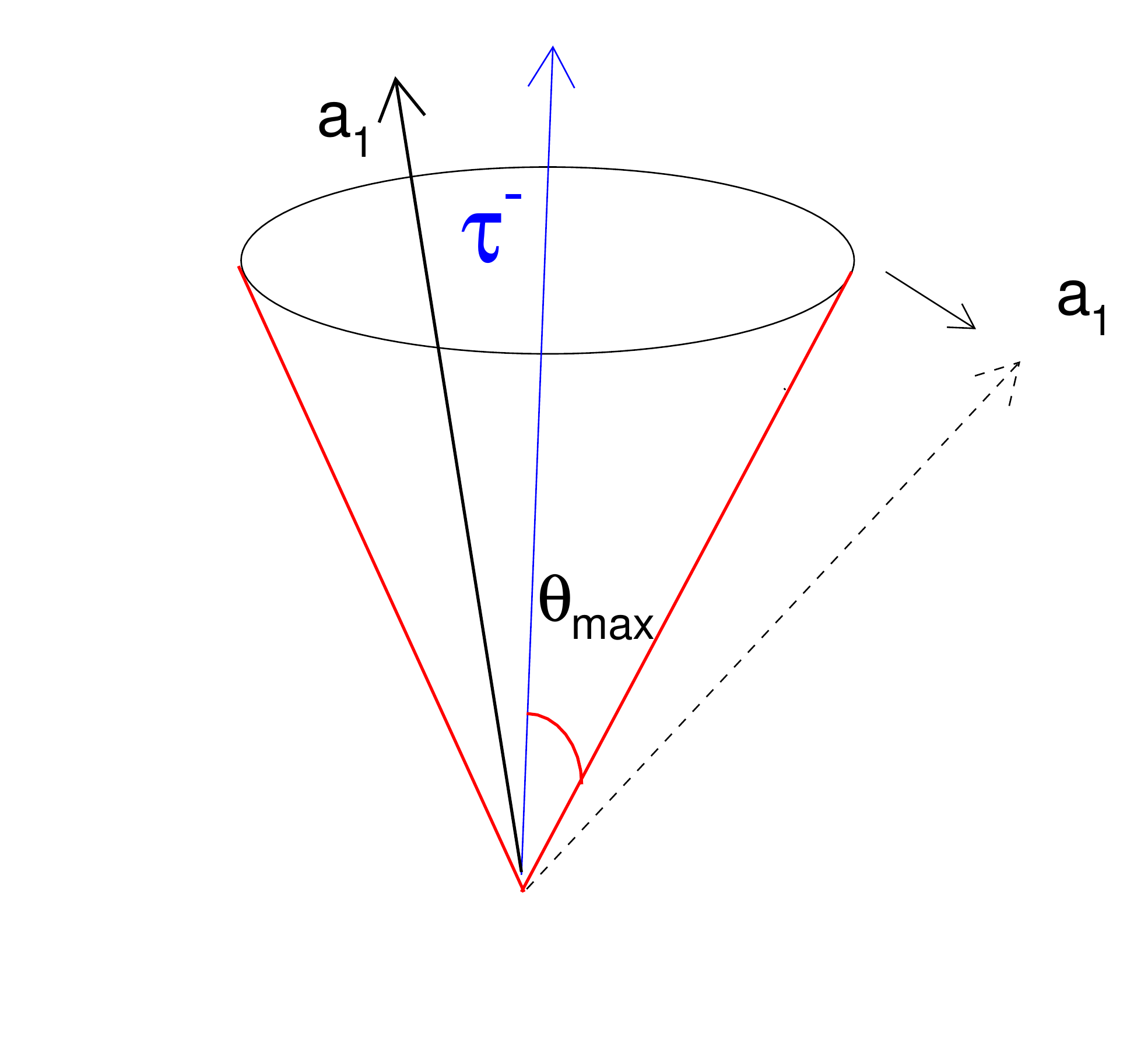}
    \caption{\textit{Left}:  Calculated tau momentum in the $\tau \rightarrow a_{1}\nu$  with: 
       $m_{a_{1}} = 1.2 ~GeV/c^{2}$ and $p_{a_{1}} = 30 ~GeV/c$. The range of allowed $\theta_{GJ}$ is limited
      by its maximal value indicated as Ambiguity point. \textit{Right}: The rotation of \ta cone in the case of unphysical  value  of $\theta_{GJ}$ in order to approach the ambiguity point.}
    \label{AmbigRota}
    \end{center}
\end{figure}
The \ta leptons produced in $Z$ decays are strongly boosted  leading to very small values of $\theta_{GJ}$.
Hence, the measured value of $\theta_{GJ}$ is very sensitive to even small shifts in the primary and secondary vertex positions. 
Without an additional quality criteria on the separation between the primary and the secondary vertex, 
a large fraction  of the decays populate the unphysical region  shown by
the shaded area in Fig.~\ref{AmbigRota} (\textit{Left}).  
Rejecting events that falls into the unphysical region or applying stronger quality criteria on the reconstruced vertices  would lead to a significant loss of statistics. 
A recovery of these events is, however, possible by moving the decay topology from the forbidden 
region to the ambiguity point as shown in Fig.~\ref{AmbigRota} (\textit{Left}). Geometrically 
this means a rotation of the \ta decay cone towards  the  $a_{1}$ direction until $\theta_{GJ}$ approaching $\theta^{max}_{GJ}$, as illustrated in 
Fig.~\ref{AmbigRota} (\textit{Right}). 
The \ta momentum in Eq.~(\ref{TauMom}) is then also shifted deteriorating  slightly the momentum resolution for these events.

Combining the direction of \ta lepton, $\vec{n}_{\tau}$  from Eq.~\ref{eqtaudir} and the absolute value of the  total momentum $|\vec{p}_{\tau}|$ from Eq.~\ref{TauMom} one can 
calculate the \ta lepton momentum $\vec{p}_{\tau} = (p^{x}_{\tau},~p^{y}_{\tau},~p^{z}_{\tau})$. If the decay falls into unphysical region, as depicted in Fig.~\ref{AmbigRota} there is only
one solution for \ta momenta, which correspond to the  maximum value of  Gottfried-Jackson angle $\theta_{GJ} = \theta_{GJ}^{max}$. If the decay is in physical region then there are two possible
values for \ta momentum that correspond to the positive and negative ambiguity points.  An iterative kinematic fit with constraints derived from the event topology allows to
resolve the ambiguity for \ta momentum as well as reconstruct the momentum of second \ta lepton. The method is described in the next section.

\section{Reconstruction of $Z \rightarrow \tau\tau \rightarrow X + 3\pi\nu$ final state.}

The reconstruction method consists in minimization of the Lagrange function constructed from the kinematic constraints with respect to
the momentum of \ta leptons. Further, by $\tau_{1}$ we denote \ta lepton that decays to three charged pions and by $\tau_{2}$ the decay $\tau \rightarrow X$.
Since the \ta leptons originate from $Z/H$ boson decays the following constraints can be introduced:

\begin{enumerate}
\item Invariant mass of the \ta leptons to be equal to  $M_{Z}/M_{H}$
\item Transverse momentum balance of both \ta leptons taking into account  a possible transverse boost of the $\tau\tau$ system.
\item Angular constraints on the $\tau$ leptons directions, derived from the measured momenta of the decay products,  positions of the SV and PV and
the helix parameters of the track from $\tau_{2}$ decay.
\end{enumerate}
\begin{figure}[hbtp]
  \begin{center}
     \includegraphics[width=0.9\textwidth]{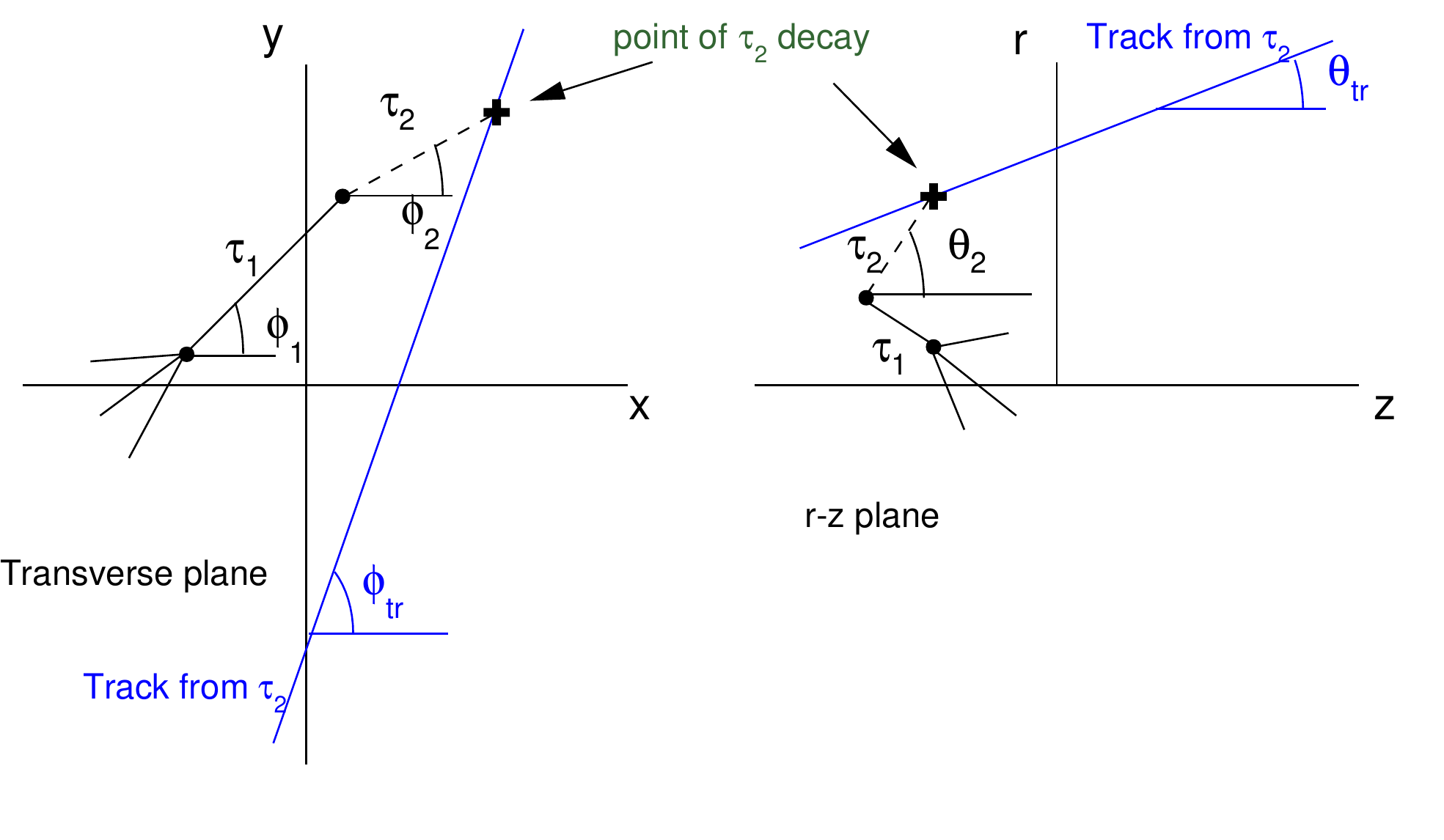}
    \caption{\textit{left} The projection of the decay  topology on the transverse plane. \textit{right} The projection of the decay  topology 
    on the rz plane. PV is the primary vertex and SV the secondary vertex of the $\tau_{1}$ decay. The angles $\phi_{1}$ and 
    $\phi_{2}$ are the azimuthal angle of the flight direction of the $\tau_{1}$ and $\tau_{2}$. $\phi_{tr}$ is the azimuthal angle of the track from $\tau_{2}$ decay.  
    $\theta_{2}$ and $\theta_{tr}$   are the polar angles of the $\tau_{2}$ and the track from $\tau_{2}$ decay. 
    The black solid cross indicates the decay point of $\tau_{2}$. }
    \label{figscetch}
  \end{center}
\end{figure}

The angular constraints are schematically shown in Fig.~\ref{figscetch}. The angle $\phi_{2}$ is estimated as
$\phi_{2} = \pi + \phi_{1} + \delta\phi$. The correction term $\delta\phi$ is introduced to account for 
the boost of the Z boson in the transverse plane and is used to estimate the starting point of the  $\tau$ leptons kinematic for the fit.   There are two alternative ways to 
estimate  $\delta\phi$:
\begin{itemize}
\item If the Z boson is boosted then the  neutrinos from \ta decays  are not flying back-to-back but tend to move in the same direction, and 
thus $\delta\phi$ can be estimated from the direction of the missing transverse energy.   \\
\end{itemize}

\begin{itemize}
\item The other approach is to use the sum of the momenta of all tracks originating from the  primary vertex, $\vec{P} = \sum\limits_{tracks} \vec{p} $ excluding ones associated with decay products of \ta leptons. 
 The direction  of $\vec{P}$ in the transverse plane strongly correlates with the transverse momentum of the Z/H boson.
\end{itemize}

Both methods are expected to have similar performance. The approach using the  missing transverse energy is explained below.

All tracks in the system to a good precision can be approximated by straight lines since the track bending radius in the magnetic field is very large
comparing to the \ta lepton flight length.

The decay point of $\tau_{2}$ is estimated as the intersection  of the  track from $\tau_{2}$ decay and the direction of $\tau_2$ given by the 
dashed line in Fig.~\ref{figscetch} (\textit{Left}). 
The corresponding z coordinate and accordingly the polar angle $\theta_{2}$ are 
determined by projecting the decay topology onto the r-z plane, as shown in Fig.~\ref{figscetch} (\textit{right}).  
The flight direction of the $\tau_{2}$ in the r-z plane is shown 
as the dashed line between the primary vertex and the point where the muon track crosses the flight direction of the $\tau_{2}$.  The angle $\theta_{2}$ is
used as an estimate for the $\tau_{2}$ flight direction in the r-z plain.

The Lagrange function is defined as follows:

\begin{equation}\label{eq:LKinFit}
\mathcal{L}(\vec{a},\vec{b}, \vec{\lambda}) =(\vec{y} - \vec{a})^{T}\pmb{V}_{y}^{-1}(\vec{y} -\vec{a}) +  \vec{f}^{T}(\vec{a},\vec{b})\pmb{V}_{f}^{-1}\vec{f}(\vec{a},\vec{b}) + 2 \vec{\lambda}^{T}\vec{H}(\vec{a},\vec{b}), 
\end{equation}
The first term  describes only the parameters of $\tau_{1}$. The vector  $\vec{y} = (p_{x},p_{y}, p_{z})$ 
comprises the measured parameters  of $\tau_{1}$ as described  in Sec.~\ref{step2} and $\pmb{V}_{y}$ is the covariance matrix 
of the parameters $\vec{y}$. 
The vectors $\vec{a}$ and $\vec{b}$ comprise the post-fit momenta of $\tau_{1}$ and $\tau_{2}$.
The  parameters for the $\tau_{1}$ and $\tau_{2}$ must match the constraints, which are separated into two parts -  the so-called soft  $\vec{f}(\vec{a},\vec{b})$  
and hard $\vec{H}(\vec{a},\vec{b})$ constraints. The  matrix $\pmb{V}_{f}$ comprises
the joint covariance matrix of parameters $\vec{a}$ and $\vec{b}$ propagated through the functions $\vec{f}$. The quantity $\vec{\lambda}$
represent the  Lagrange multipliers.

The momentum of the $\tau_{1}$ and the $\tau_{2}$,  are obtained by minimizing $\mathcal{L}$ from the Eq.~(\ref{eq:LKinFit}) with respect to vectors $\vec{a}$, $\vec{b}$ and   $\vec{\lambda}$ as described in Sec.~\ref{EFformalism}.

The constraints on the invariant mass of the di-\ta system and on the longitudinal 
momentum component of the $\tau_{2}$ are introduced as hard constraints $\vec{H} =0$, i.e.:

\begin{equation}\label{hardconstr}
  \vec{H} = \left \{
  \begin{array}{l l}
    M_{\tau\tau} - M_{Z/H} &  \\
    p_{z} - |\vec{p}_{2}|\cos\theta_{c} & \\
   \end{array} \right.   \hspace{1cm}
\end{equation}

 where $M_{Z}$ is the  mass of the Z or H boson, $M_{\tau\tau}$ the invariant mass of di-tau system, $p_{z}$ and $|\vec{p}_{2}|$ are the longitudinal 
component and the total momentum of $\tau_2$. There are two options for the constraint on the polar angle $\theta_{c}$ of  $\tau_{2}$: 

\begin{enumerate}
\item  $\theta_{c}$  =  $\theta_{2}$  as described in  Fig.~\ref{figscetch} (\textit{ Right })
\end{enumerate}

\begin{enumerate}
  \setcounter{enumi}{1}
\item  $\theta_{c} \simeq  \theta_{tr}$, this  implies that the $\tau$ leptons in the decay of the $Z/H$ boson are highly boosted and consequently 
the decay products are collimated in a small cone around the $\tau$ lepton flight direction. This approach is also called a \textit{collinear approximation}. This  should be noted that
the angle between the \ta decay products  and the \ta lepton carries the information on the helicity state of the \ta lepton  which is smash out in the collinear approximation approach.
\end{enumerate}

In order to account for the transverse boost of the $Z/H$ boson, the soft constraint term  $\vec{f}$ is defined as:

\begin{equation}\label{softconstr}
 \vec{f} = \left \{
  \begin{array}{l l}
    p^{\tau_{1}}_{x} + p^{\tau_{2}}_{x} - p^{a_{1}}_{x} - p^{vis_{2}}_{x} - MET_{x} & \\
    p^{\tau_{1}}_{y} + p^{\tau_{2}}_{y} - p^{a_{1}}_{y} - p^{vis_{2}}_{y} - MET_{y} & \\
   \end{array} \right\}
\end{equation}

Here $\vec{p}^{\tau_{1}}$, $\vec{p}^{\tau_{2}}$, $\vec{p}^{a_{1}}$, $\vec{p}^{vis_{2}}$ are the momenta of the \ta leptons, the  $a_1$ and 
the visible decay products from the decay of $\tau_{2}$, respectively, and
$MET_{x}$ and $MET_{y}$ are the x and y components of the missing transverse energy in the event (the inverted sum of momentum of  all detected particles in the event).
In total there are three unknown parameters, the momentum of $\tau_{2}$,  $(p^{x}_{\tau_{2}},~p^{y}_{\tau_{2}},~p^{z}_{\tau_{2}})$ with four constraints.  The ambiguity of momentum of $\tau_{1}$ can
be resolved by performing the minimization of Lagrangian~(\ref{eq:LKinFit}) two times with vector $\vec{y} = \vec{y}_{\pm}$ corresponding to the positive and negative ambiguity point for $\tau_{1}$  momentum.  
The decision on which  ambiguity point  should be given a preference can be taken by comparing the minimized value of Lagrangian for positive and negative ambiguity points, $\mathcal{L}_{\pm}(\vec{a},\vec{b}, \vec{\lambda})$.
In case of the truth momentum of $\tau_{1}$ corresponds to the positive ambiguity point then $\mathcal{L}^{min}_{+}(\vec{a},\vec{b}, \vec{\lambda}) < \mathcal{L}^{min}_{-}(\vec{a},\vec{b}, \vec{\lambda})$ and vice versa in case of the negative ambiguity point.

If both \ta leptons decay into $a_{1}$ resonance and three charged pions then the kinematic of both \ta leptons can be calculated as described in Sec.~\ref{step2}. The angular constraint in~(\ref{hardconstr} can be
dropped and one can keep only the mass term. The Lagrange function to be minimized is modified to: 

\begin{equation}\label{eq:LKinFit2}
\begin{split}
\mathcal{L}(\vec{a},\vec{b}, \vec{\lambda}) =(\vec{y} - \vec{a})^{T}\pmb{V}_{y}^{-1}(\vec{y} -\vec{a}) + (\vec{z} - \vec{b})^{T}\pmb{V}_{z}^{-1}(\vec{z} -\vec{b}) \\
+  \vec{f}^{T}(\vec{a},\vec{b})\pmb{V}_{f}^{-1}\vec{f}(\vec{a},\vec{b}) + 2 \vec{\lambda}^{T}\vec{H}(\vec{a},\vec{b}), 
\end{split}
\end{equation}
where the additional $\chi^{2}$ term is included which   describes only the parameters of $\tau_{2}$. The vector  $\vec{z} = (p_{x},p_{y}, p_{z})$ 
comprises the measured parameters  of $\tau_{2}$ as described  in Sec.~\ref{step2} and $\pmb{V}_{z}$ is the covariance matrix 
of the parameters $\vec{z}$.  In this case the momentum of both \ta leptons are known with ambiguities and in order to figure out the correct kinematic one needs to study four combinations
$\mathcal{L}^{min}_{+-}(\vec{a},\vec{b}, \vec{\lambda})$, $\mathcal{L}^{min}_{++}(\vec{a},\vec{b}, \vec{\lambda})$, $\mathcal{L}^{min}_{--}(\vec{a},\vec{b}, \vec{\lambda})$ and $\mathcal{L}^{min}_{-+}(\vec{a},\vec{b}, \vec{\lambda})$, where signs denote the positive or negative ambiguity point for momentum of the first and second \ta lepton ($\vec{y} = \vec{y}_{\pm}$, $\vec{z} = \vec{z}_{\pm}$). The smallest of these minimized Lagrangians indicates the corresponding momentum ambiguity point for both \ta leptons.

\section{Event Fit Formalism} \label{EFformalism}
The method of Lagrange multipliers allows to minimize or maximize a function with  constraints. In this section we describe a general solution for the unknown parameters $\vec{a}$ and $\vec{b}$ with the set of constraints $\vec{H}=0$, $\vec{f} =0$ and the set of measured parameters  $\vec{y}$. The Lagrange function to be minimized is:
\begin{equation}\label{eq:L}
\mathcal{L}(\vec{a}, \vec{b}, \vec{\lambda}) =(\vec{y} - \vec{a})^{T}\pmb{V}_{y}^{-1}(\vec{y} -\vec{a}) +  \vec{f}^{T}(\vec{a},\vec{b})\pmb{V}_{f}^{-1}\vec{f}(\vec{a},\vec{b}) + 2 \vec{\lambda}^{T}\vec{H}(\vec{a},\vec{b}) 
\end{equation}
The vectors $\vec{a}$ and $\vec{b}$ are composed of  $n$ and $m$ parameters:
\begin{equation}
\vec{a} = 
\begin{pmatrix}
  a_{1} \\
  a_{2} \\
  \vdots \\
  a_{n}
\end{pmatrix} \hspace{0.8cm}
\vec{b} = 
\begin{pmatrix}
  b_{1} \\
  b_{2} \\
  \vdots \\
  b_{m}.
\end{pmatrix}
\end{equation}

$\pmb{V}_{y}$ is the covariance matrix of parameters $\vec{a}$ and $\pmb{V}_{f}$ is a covariance matrix of both $\vec{a}$ and $\vec{b}$ propagated through the functions $\vec{f}$.

The $r$ functions $\vec{H}$ and $k$ functions $\vec{f}$, describing  physical constraints, are defined in a form:
\begin{equation}\label{eq:constr}
 \vec{H}(\vec{a}, \vec{b}) =0   \hspace{0.8cm} \vec{f}(\vec{a}, \vec{b}) = 0.
\end{equation}

It is assumed that the constraint functions can be linearized by expanding around some convenient points $\vec{a}_{0}$ and  $\vec{b}_{0}$.  
In the iterative procedure the points of expansion for every iteration can be chosen as the solution for $\vec{a}$ and $\vec{b}$ obtained in the previous iteration. 
While for the first iteration a reasonable,  estimation of the parameters of interest  serves as  expansion points. 

Expanding Eq.~(\ref{eq:constr}) gives the linearized equations:
\begin{equation}\label{eq:linconstr}
\begin{split}
 \pmb{H}_{a}\Delta\vec{a}  + \pmb{H}_{b}\Delta\vec{b}  + \vec{H}_{0}(\vec{a}_{0},\vec{b}_{0}) =0\\
 \pmb{F}_{a}\Delta\vec{a}  + \pmb{F}_{b}\Delta\vec{b}  + \vec{F}_{0}(\vec{a}_{0},\vec{b}_{0}) =0\\
\end{split}
\end{equation}

where $\Delta\vec{a} = \vec{a} - \vec{a}_{0}$,  $\Delta\vec{b} = \vec{b} - \vec{b}_{0}$. The matrices $\pmb{H}_{a}$, $\pmb{H}_{b}$, $\pmb{F}_{a}$, $\pmb{F}_{b}$ are Jacobians of the constraints $\vec{H}$ and $\vec{f}$
with respect to parameters $\vec{a}$ and $\vec{b}$ and given by:

\begin{equation}
\pmb{H}_{a} = 
\begin{pmatrix}
   \frac{\partial H_{1}}{\partial a_{1}} & \frac{\partial H_{1}}{\partial a_{2}} & \hdots & \frac{\partial H_{1}}{\partial a_{n}}\\
   \frac{\partial H_{2}}{\partial a_{1}} & \frac{\partial H_{2}}{\partial a_{2}} & \hdots & \frac{\partial H_{2}}{\partial a_{n}}\\
   \vdots & \vdots &  \ddots & \vdots   \\
   \frac{\partial H_{r}}{\partial a_{1}} & \frac{\partial H_{r}}{\partial a_{2}} & \hdots & \frac{\partial H_{r}}{\partial a_{n}}
\end{pmatrix} \hspace{0.8cm}
\pmb{H}_{b} = 
\begin{pmatrix}
   \frac{\partial H_{1}}{\partial b_{1}} & \frac{\partial H_{1}}{\partial b_{2}} & \hdots & \frac{\partial H_{1}}{\partial b_{m}}\\
   \frac{\partial H_{2}}{\partial b_{1}} & \frac{\partial H_{2}}{\partial b_{2}} & \hdots & \frac{\partial H_{2}}{\partial b_{m}}\\
   \vdots & \vdots &  \ddots & \vdots   \\
   \frac{\partial H_{r}}{\partial b_{1}} & \frac{\partial H_{r}}{\partial b_{2}} & \hdots & \frac{\partial H_{r}}{\partial b_{m}}
\end{pmatrix}
\end{equation}
\begin{equation}
\pmb{F}_{a} = 
\begin{pmatrix}
   \frac{\partial F_{1}}{\partial a_{1}} & \frac{\partial F_{1}}{\partial a_{2}} & \hdots & \frac{\partial F_{1}}{\partial a_{n}}\\
   \frac{\partial F_{2}}{\partial a_{1}} & \frac{\partial F_{2}}{\partial a_{2}} & \hdots & \frac{\partial F_{2}}{\partial a_{n}}\\
   \vdots & \vdots &  \ddots & \vdots   \\
   \frac{\partial F_{k}}{\partial a_{1}} & \frac{\partial F_{k}}{\partial a_{2}} & \hdots & \frac{\partial F_{k}}{\partial a_{n}}
\end{pmatrix} \hspace{0.8cm}
\pmb{F}_{b} = 
\begin{pmatrix}
   \frac{\partial F_{1}}{\partial b_{1}} & \frac{\partial F_{1}}{\partial b_{2}} & \hdots & \frac{\partial F_{1}}{\partial b_{m}}\\
   \frac{\partial F_{2}}{\partial b_{1}} & \frac{\partial F_{2}}{\partial b_{2}} & \hdots & \frac{\partial F_{2}}{\partial b_{m}}\\
   \vdots & \vdots &  \ddots & \vdots   \\
   \frac{\partial F_{k}}{\partial b_{1}} & \frac{\partial F_{k}}{\partial b_{2}} & \hdots & \frac{\partial F_{k}}{\partial b_{m}}
\end{pmatrix}
\end{equation}

$\vec{H}_{0}$ and $\vec{F}_{0}$ are the constraint values at the linearization point:

\begin{equation}
\vec{H}_{0} = 
\begin{pmatrix}
    H_{1}(\vec{a}_{0},\vec{b}_{0})\\
    H_{2}(\vec{a}_{0},\vec{b}_{0})\\
    \vdots   \\
    H_{r}(\vec{a}_{0},\vec{b}_{0})
\end{pmatrix}\hspace{0.8cm}
\vec{F}_{0} = 
\begin{pmatrix}
    F_{1}(\vec{a}_{0},\vec{b}_{0})\\
    F_{2}(\vec{a}_{0},\vec{b}_{0})\\
    \vdots   \\
    F_{k}(\vec{a}_{0},\vec{b}_{0}).
\end{pmatrix}
\end{equation}

It is worth noting, that the actual representation of the parameters $\vec{a}$ and  $\vec{b}$ must be  chosen such that the constraint functions are linear. However, often this is not possible which
leads to the fact that the Eq.~(\ref{eq:linconstr}) are not longer acceptable approximations. In case of  highly non-linear constraints one can diminish this effect by expanding Eq.~(\ref{eq:constr}) to the second or higher orders, but it will
make further calculations much more complex.

The solution for set of parameters  $\vec{a}$ and $\vec{b}$ can be found now by minimizing Eq.~(\ref{eq:L}) with respect to $\vec{a}$, $\vec{b}$ and $\vec{\lambda}$. 
Substituting  in Eq.~(\ref{eq:L}) $\vec{H}$ and $\vec{f}$  by:

\begin{equation}
\begin{split}
\vec{H}(\vec{a},\vec{b}) &= \pmb{H}_{a}(\vec{a} - \vec{a}_{0}) +    \pmb{H}_{b}(\vec{b} - \vec{b}_{0})  + \vec{H}_{0}(\vec{a}_{0},\vec{b}_{0}) \\
\vec{f}(\vec{a},\vec{b}) &= \pmb{F}_{a}(\vec{a} - \vec{a}_{0}) +    \pmb{F}_{b}(\vec{b} - \vec{b}_{0})  + \vec{F}_{0}(\vec{a}_{0},\vec{b}_{0}) 
\end{split}
\end{equation}

and performing derivation one gets the equations:

\begin{equation}
\scriptsize
\begin{split}
\frac{\partial \mathcal{L}(\vec{a}, \vec{b}, \vec{\lambda})}{\partial \vec{a}} &= -\pmb{V}_{y}^{-1}\vec{y} + \pmb{V}^{-1}\vec{a} + \pmb{F}^{T}_{a}\pmb{V}^{-1}_{f}\vec{F}_{0} + \pmb{F}^{T}_{a}\pmb{V}^{-1}_{f}\pmb{F}^{T}_{a}\Delta\vec{a}  + \pmb{F}^{T}_{a}\pmb{V}^{-1}_{f}\pmb{F}^{T}_{b}\Delta\vec{b} + \pmb{H}^{T}_{a}\vec{\lambda} =0\\
\frac{\partial \mathcal{L}(\vec{a}, \vec{b}, \vec{\lambda})}{\partial \vec{b}} &=  \pmb{F}^{T}_{b}\pmb{V}^{-1}_{f}\vec{F}_{0} + \pmb{F}^{T}_{b}\pmb{V}^{-1}_{f}\pmb{F}^{T}_{a}\Delta\vec{a} + \pmb{F}^{T}_{b}\pmb{V}^{-1}_{f}\pmb{F}^{T}_{b}\Delta\vec{b}  + \pmb{H}_{b}^{T}\vec{\lambda} =0 \\
\frac{\partial \mathcal{L}(\vec{a}, \vec{b}, \vec{\lambda})}{\partial \vec{\lambda}} &= \pmb{H}_{a}\Delta\vec{a} + \pmb{H}_{b}\Delta\vec{b}  + \vec{H}_{0}  =0.
\end{split}
\end{equation}

It is straightforward to find the parameters $\vec{a}$, $\vec{b}$ and $\vec{\lambda}$ from these three linear equations, after some algebra one can write them in a short matrix form as:

\begin{equation}\label{solution}
\scriptsize
 \begin{pmatrix}
  \pmb{V}_{y}^{-1} + \pmb{F}^{T}_a\pmb{V}^{-1}_{f}\pmb{F}_{a} &   & \pmb{F}^{T}_a\pmb{V}^{-1}_{f}\pmb{F}_{b} &   & \pmb{H}^{T}_{a}\\
                             &   &                     &   &    \\
  \pmb{F}^{T}_b\pmb{V}^{-1}_{f}\pmb{F}_{a} &   & \pmb{F}^{T}_b\pmb{V}^{-1}_{f}\pmb{F}_{b} &   & \pmb{H}^{T}_{b} \\
                             &   &                     &   &    \\
  \pmb{H}_{a} &   & \pmb{H}_{b} &   &  \pmb{0}
 \end{pmatrix} \begin{pmatrix} 
\vec{a}\\
 \\
\vec{b}\\
 \\
\vec{\lambda}\\
  \end{pmatrix}= \begin{pmatrix}
\pmb{V}_{y}^{-1}\vec{y} - \pmb{F}^{T}_{a}\pmb{V}^{-1}_{f}(\vec{F}_{0} - \pmb{F}_{a}\vec{a}_{0} - \pmb{F}_{b}\vec{b}_{0})\\
 \\
-\pmb{F}^{T}_{b}\pmb{V}^{-1}_{f}(\vec{F}_{0} - \pmb{F}_{a}\vec{a}_{0} - \pmb{F}_{b}\vec{b}_{0})\\
 \\
\pmb{H}_{a}\vec{a}_{0} + \pmb{H}_{b}\vec{b}_{0} - \vec{H}_{0}\\
 \end{pmatrix}
\end{equation}

In case of both \ta leptons decay to three charged pions and with the modified Lagrange function~(\ref{eq:LKinFit2}) the solution for paramters $\vec{a}$, $\vec{b}$ and $\vec{\lambda}$ is:
\begin{equation}\label{solutiona1a1}
\scriptsize
 \begin{pmatrix}
  \pmb{V}_{y}^{-1} + \pmb{F}^{T}_a\pmb{V}^{-1}_{f}\pmb{F}_{a} &   & \pmb{F}^{T}_a\pmb{V}^{-1}_{f}\pmb{F}_{b} &   & \pmb{H}^{T}_{a}\\
                             &   &                     &   &    \\
  \pmb{F}^{T}_b\pmb{V}^{-1}_{f}\pmb{F}_{a} &   & \pmb{V}_{z}^{-1} + \pmb{F}^{T}_b\pmb{V}^{-1}_{f}\pmb{F}_{b} &   & \pmb{H}^{T}_{b} \\
                             &   &                     &   &    \\
  \pmb{H}_{a} &   & \pmb{H}_{b} &   &  \pmb{0}
 \end{pmatrix} \begin{pmatrix} 
\vec{a}\\
 \\
\vec{b}\\
 \\
\vec{\lambda}\\
  \end{pmatrix}= \begin{pmatrix}
\pmb{V}_{y}^{-1}\vec{y} - \pmb{F}^{T}_{a}\pmb{V}^{-1}_{f}(\vec{F}_{0} - \pmb{F}_{a}\vec{a}_{0} - \pmb{F}_{b}\vec{b}_{0})\\
 \\
\pmb{V}_{z}^{-1}\vec{z} -\pmb{F}^{T}_{b}\pmb{V}^{-1}_{f}(\vec{F}_{0} - \pmb{F}_{a}\vec{a}_{0} - \pmb{F}_{b}\vec{b}_{0})\\
 \\
\pmb{H}_{a}\vec{a}_{0} + \pmb{H}_{b}\vec{b}_{0} - \vec{H}_{0}\\
 \end{pmatrix}
\end{equation}

The corresponding value of $\mathcal{L}_{min}(\vec{a}, \vec{b}, \vec{\lambda})$ can be found by evaluating  Eq.~(\ref{eq:L}) with the  parameters $\vec{a}$, $\vec{b}$, $\vec{\lambda}$ obtained from Eq.~(\ref{solution}).
The iterative approach to the problem consist in substituting the linearization point $\vec{a}_{0}$ and $\vec{b}_{0}$ by $\vec{a}$, $\vec{b}$ obtained from the previous step and repeating the procedure until some convergency requirements are fulfilled. As a convergency requirement one can use the value of parameters change  after every iteration $\Delta \vec{a} = \vec{a}_{i} - \vec{a}_{i-1}$ and $\Delta \vec{b} = \vec{b}_{i} - \vec{b}_{i-1}$, the iterations can be stopped if $\Delta \vec{a}$ and $\Delta \vec{a}$ are not large with respect to the first iteration. The other possible convergency requirement can be the change of $\mathcal{L}$ in the last iteration, $\Delta \mathcal{L} = \mathcal{L}_{i} - \mathcal{L}_{i-1}$.  The case $\Delta \mathcal{L} =0$ indicates that the Lagrange function reaches its minimum $\mathcal{L}_{i} = \mathcal{L}^{min}$ at iteration $i$.

\section{Summary}\label{summarysection}
The Global Event Fit method is an experimental technique for reconstruction of full kinematic of the $\tau\tau$ pair originated from the decay of $Z$ or $H$ bosons.
The advantage of this method is that it is based mainly on the measured experimental input with only one assumption on the mass of the origin resonance. The method also 
avoids the limitation of the collinear approximation and can be applied to any $\tau\tau$ decays when on of the \ta decays into three charged pions.  The described Global Event Fit method
demonstrated a good performance in measurement of the $\tau$ lepton helicity separation~\cite{CMS-DP-2016-060} with the CMS detector~\cite{Chatrchyan:2008aa}.

\bibliographystyle{ieeetr}
\bibliography{gefBiblio}

\end{document}